\documentstyle[twoside,epsf]{article}

\catcode`\@=11
\long\def\@makefntext#1{
\protect\noindent \hbox to 3.2pt {\hskip-.9pt  
$^{{\eightrm\@thefnmark}}$\hfil}#1\hfill}		

\def\@makefnmark{\hbox to 0pt{$^{\@thefnmark}$\hss}}	
	
\def\ps@myheadings{\let\@mkboth\@gobbletwo
\def\@oddhead{\hbox{}
\rightmark\hfil\eightrm\thepage}   
\def\@oddfoot{}\def\@evenhead{\eightrm\thepage\hfil
\leftmark\hbox{}}\def\@evenfoot{}
\def\sectionmark##1{}\def\subsectionmark##1{}}

\oddsidemargin=\evensidemargin
\newcounter{sectionc}\newcounter{subsectionc}\newcounter{subsubsectionc}
\renewcommand{\section}[1] {\vspace{12pt}\addtocounter{sectionc}{1} 
\setcounter{subsectionc}{0}\setcounter{subsubsectionc}{0}\noindent 
	{\tenbf\thesectionc. #1}\par\vspace{5pt}}
\renewcommand{\subsection}[1] {\vspace{12pt}\addtocounter{subsectionc}{1} 
	\setcounter{subsubsectionc}{0}\noindent 
	{\bf\thesectionc.\thesubsectionc. {\kern1pt \bfit #1}}\par\vspace{5pt}}
\renewcommand{\subsubsection}[1] {\vspace{12pt}\addtocounter{subsubsectionc}{1}
	\noindent{\tenrm\thesectionc.\thesubsectionc.\thesubsubsectionc.
	{\kern1pt \tenit #1}}\par\vspace{5pt}}
\newcommand{\nonumsection}[1] {\vspace{12pt}\noindent{\tenbf #1}
	\par\vspace{5pt}}

\newcommand{\textlineskip}{\baselineskip=13pt}
\newcommand{\smalllineskip}{\baselineskip=10pt}

\def\eightcirc{
\begin{picture}(0,0)
\put(4.4,1.8){\circle{6.5}}
\end{picture}}
\def\eightcopyright{\eightcirc\kern2.7pt\hbox{\eightrm c}} 

\newcommand{\publisher}[2]{{\begin{center}\footnotesize\smalllineskip 
	Received #1\\
	Revised #2
	\end{center}
	}}

\def\abstracts#1#2#3{{
	\centering{\begin{minipage}{4.5in}\baselineskip=10pt\footnotesize
	\parindent=0pt #1\par 
	\parindent=15pt #2\par
	\parindent=15pt #3
	\end{minipage}}\par}} 

\renewenvironment{thebibliography}[1]
	{\frenchspacing
	 \ninerm\baselineskip=11pt
	 \begin{list}{\arabic{enumi}.}
        {\usecounter{enumi}\setlength{\parsep}{0pt}     
	 \setlength{\leftmargin 12.7pt}{\rightmargin 0pt} 
         \setlength{\itemsep}{0pt} \settowidth
	{\labelwidth}{#1.}\sloppy}}{\end{list}}

\newcounter{itemlistc}
\newcounter{romanlistc}
\newcounter{alphlistc}
\newcounter{arabiclistc}

\def\@citex[#1]#2{\if@filesw\immediate\write\@auxout
	{\string\citation{#2}}\fi
\def\@citea{}\@cite{\@for\@citeb:=#2\do
	{\@citea\def\@citea{,}\@ifundefined
	{b@\@citeb}{{\bf ?}\@warning
	{Citation `\@citeb' on page \thepage \space undefined}}
	{\csname b@\@citeb\endcsname}}}{#1}}

\newif\if@cghi
\def\cite{\@cghitrue\@ifnextchar [{\@tempswatrue
	\@citex}{\@tempswafalse\@citex[]}}
\def\citelow{\@cghifalse\@ifnextchar [{\@tempswatrue
	\@citex}{\@tempswafalse\@citex[]}}
\def\@cite#1#2{{$\null^{#1}$\if@tempswa\typeout
	{IJCGA warning: optional citation argument 
	ignored: `#2'} \fi}}

\def\@refcitex[#1]#2{\if@filesw\immediate\write\@auxout
	{\string\citation{#2}}\fi
\def\@citea{}\@refcite{\@for\@citeb:=#2\do
	{\@citea\def\@citea{, }\@ifundefined
	{b@\@citeb}{{\bf ?}\@warning
	{Citation `\@citeb' on page \thepage \space undefined}}
	\hbox{\csname b@\@citeb\endcsname}}}{#1}}

\def\@refcite#1#2{{#1\if@tempswa\typeout
        {IJCGA warning: optional citation argument
	ignored: `#2'} \fi}}

\def\refcite{\@ifnextchar[{\@tempswatrue
	\@refcitex}{\@tempswafalse\@refcitex[]}}

\def\pmb#1{\setbox0=\hbox{#1}
	\kern-.025em\copy0\kern-\wd0
	\kern.05em\copy0\kern-\wd0
	\kern-.025em\raise.0433em\box0}

\def\fnt#1#2{\footnotetext{\kern-.3em
	{$^{\mbox{\scriptsize #1}}$}{#2}}}

\def\runninghead#1#2{\pagestyle{myheadings}
\markboth{{\protect\footnotesize\it{\quad #1}}\hfill}
{\hfill{\protect\footnotesize\it{#2\quad}}}}
\font\tenrm=cmr10
\font\tenit=cmti10 
\font\tenbf=cmbx10
\font\bfit=cmbxti10 at 10pt
\font\ninerm=cmr9
\font\nineit=cmti9
\font\ninebf=cmbx9
\font\eightrm=cmr8

\font\sevenrm=cmr7

\def\qed{\hbox{${\vcenter{\vbox{			
   \hrule height 0.4pt\hbox{\vrule width 0.4pt height 6pt
   \kern5pt\vrule width 0.4pt}\hrule height 0.4pt}}}$}}

\begin{document}

\newpage

\runninghead{Haret C. Rosu} 
{Darboux cosmological fluids}

\normalsize\textlineskip
\thispagestyle{empty}
\setcounter{page}{1}


\vspace*{0.88truein}

\bigskip
\centerline{\bf  DARBOUX CLASS OF COSMOLOGICAL FLUIDS WITH} 
\centerline{{\bf TIME-DEPENDENT ADIABATIC INDICES} 
\footnote{
This essay received an ``honorable mention'' in the 
        Annual Essay Competition of the Gravity Research 
        Foundation for the year 2000 - Ed.}}
\vspace*{0.035truein}
\vspace*{0.37truein}
\vspace*{10pt}
\centerline{\footnotesize H.C. ROSU}
\vspace*{0.015truein}
\centerline{\footnotesize  Instituto de F\'{\i}sica,
Universidad de Guanajuato, Apdo Postal E-143, 37150 Le\'on, Gto, Mexico}
\centerline{\footnotesize International Center for Relativistic Astrophysics, 
Rome-Pescara, Italy}
\vspace*{0.225truein}
\publisher{(16 May 2000) - Mod Phys. Lett. A 15 (May 2000) 979-990}

\vspace*{0.21truein}
\abstracts{A one-parameter family of
time dependent adiabatic indices is introduced for
any given type of cosmological fluid of constant adiabatic index
by a mathematical method belonging to the class
of Darboux transformations. The procedure works for zero cosmological
constant at the price of introducing a new constant parameter
related to the time dependence of the adiabatic index.
These fluids can be the real cosmological fluids that are encountered
at cosmological scales and they could be used as a simple and efficient
explanation for the recent experimental findings regarding the present day
accelerating universe. In addition, new types of cosmological scale 
factors, corresponding to these fluids, are presented.}{}{}


\textlineskip                  
\vspace*{12pt}                 

\vspace*{1pt}\textlineskip	
\vspace*{-0.5pt}
\noindent


\noindent





{\bf 1} - Recent data from type Ia supernovae
that can still be considered as meager provoked tremendous interest
since they indicate
that the universe has {\em now} an accelerating 
large scale expansion,\cite{type1a} (${\rm \rho+3p<0}$ in Eq.~(1) below).
This supernovae revolution stimulated
model builders to supply a remarkable amount of paper work on this new
topic, in which such impressive words as {\em quintessence} can be found.
On the other hand,
simple explanations should always be taken into account on the background
of ever more complicated modeling.
In the following, I would like to act according to the latter rule
and present a rather nontrivial contribution that can seemingly be related 
to accelerating universes.

\bigskip

{\bf 2} - The scale factor ${\rm a(t)}$ of a FLRW metric is a function of the
comoving time ${\rm t}$ obeying the Einstein-Friedmann
dynamical equations (EFDEs):

$$
{\rm \frac{\ddot{a}}{a}=-\frac{4\pi G}{3}(\rho +3p)}
\eqno(1)
$$
$$
{\rm
\left(\frac{\dot{a}}{a}\right)^2=\frac{8\pi G\rho}{3}-\frac{\kappa}{a^2}~,}
\eqno(2)
$$
where $\rho$ and ${\rm p}$ are the energy
density and the pressure, respectively,
of the perfect fluid of which a classical universe is usually assumed to be
made of, and $\kappa=0,\pm1$ is the curvature index of the flat, closed, open
universe, respectively.
To solve EFDEs for ${\rm a(t)}$ one needs an equation of state for the
fluid, i.e., a relationship between ${\rm p}$ and $\rho$. Very used
ones are the barotropic equations of state (BES)
$$
{\rm p=(\gamma -1)\rho,}
\eqno(3)
$$
where $\gamma$ is the constant adiabatic index. As is well known, all the
important cosmological fluids assumed to be dominant along various
cosmological epochs belong to the barotropic class.
It is straightforward to get ${\rm a(t)}$ by direct integration
for the flat case, however there are
some difficulties in the other two cases. Landau and Lifschitz \cite{LL}
gave a general procedure to perform the integration, which is valid for
any $\gamma$. On the other hand, for the purposes of the present work,
an equivalent procedure introduced by Faraoni \cite{Far} leading
practically to the same results (however with key constraints)
is very useful.

\bigskip

{\bf 3} - The alternative method of Faraoni
to derive the scale factor of the universe is based on the Riccati equation
and goes as follows. First, one combines EFDEs and BES yielding
$$
{\rm
\frac{\ddot{a}}{a}+c\left(\frac{\dot{a}}{a}\right)^2+\frac{c\kappa}{a^2}=0~,}
\eqno(4)
$$
where ${\rm c=\frac{3}{2}\gamma -1}$. The case $\kappa=0$ is directly
integrable \cite{Far}
and I shall not pay attention to it in the present work though it can be
included in the scheme.
Eq.~(4) is rewritten in the conformal time variable $\eta$ in the
form
$$
{\rm \frac{a^{''}}{a}+(c-1)\left(\frac{a^{'}}{a}\right)^2+c\kappa=0~.}
\eqno(5)
$$
One can immediately see that by means of the change of function
$u=\frac{a^{'}}{a}$ the following Riccati equation is obtained
(${\rm c > 0}$ henceforth)
$$
{\rm u^{'}+cu^2+\kappa c=0~.}
\eqno(6)
$$

Furthermore, employing ${\rm u=\frac{1}{c}\frac{w^{'}}{w}}$ one gets the
very simple second order differential equation
$$
{\rm w^{''}+\kappa c^2w=0~.}
\eqno(7a)
$$
For $\kappa =1$ the solution of the latter is
${\rm w_{1}=W_1\cos(c\eta +d)}$ , where ${\rm d}$ is an arbitrary phase,
implying
$$
{\rm a_{1}(\eta)=A_1[\cos(c\eta +d)]^{1/c}~,}
\eqno(8)
$$
whereas for $\kappa =-1$ one gets
${\rm w_{-1}=W_{-1}{\rm sinh}(c\eta)}$ and therefore
$$
{\rm a_{-1}(\eta)=A_{-1}[{\rm sinh}(c\eta)]^{1/c}}~,
\eqno(9)
$$
where ${\rm W_{\pm 1}}$ and ${\rm A_{\pm 1}}$ are amplitude parameters.
Eqs. (8) and (9) are the same solutions as in the standard procedure.
According to Faraoni, the alternative method requires $\gamma$ be a
constant. However, as I will show
in the following there is still some room in this method
for time-dependent adiabatic indices.

\bigskip

{\bf 4} - I will now apply the so-called strictly isospectral
Darboux technique (SIDT\footnote{In my previous works,I have also used
the acronym DDGR for double Darboux general Riccati to denote the
same scheme.)})
to Eq. (7a). This is a mathematical approach related to the
factorizations of the second order linear differential equations that
underlined supersymmetric quantum mechanics and the dynamical symmetry
breaking ideas of Witten almost two decades ago.\cite{W81} There is a
slight but important difference between Witten's approach and SIDT.
While Witten's factorizations rely on the
particular Riccati solution, the SIDT uses the general Riccati solution.
This was first discussed by Mielnik in 1984. \cite{M84} The difference with
respect to Mielnik is that I will employ the SIDT scheme at fixed {\em zero}
`energy' (see also \cite{PSP}).
I refer the reader to my
short survey of Darboux transformations for more details.\cite{Ro}
During the last few years, I have acquired some experience with SIDT
extending its range of applications beyond supersymmetric quantum
mechanics, to subjects such as Newtonian damping, \cite{N}
Fokker-Planck equation,\cite{FP}
and quantum cosmology.\cite{QC}
It is my aim here to apply this mathematical
scheme to the dynamics of the scale factor of a classical universe.

First of all, despite ${\rm c}$ looks like a frequency for the function
${\rm w}$, I shall interpret
${\rm \pm c}$ as Schr\"odinger `potentials'.
This is preferable to ${\rm \pm c^2}$ for
algebraic reasons coming out from the bunch of formulas that I present next.
(Notice that we have a constant potential having a coupling constant equal
to the potential itself!).
Thus, Eq.~(7a) will be considered as a Schr\"odinger
equation at zero `energy' (or more rigorously at zero factorization constant).

The point now is that the Riccati solution
${\rm u_{p}=\frac{1}{c}\frac{w^{'}}{w}}$
mentioned above is only the particular solution, i.e.,
${\rm u_{p,1}=-\tan c\eta}$ and ${\rm u_{p,-1}={\rm coth} c\eta}$
for $\kappa =\pm 1$, respectively. The particular Riccati solutions are
closely related to the common factorizations of the Schr\"odinger equation.
Indeed, Eq.~(7a) can be written
$$
{\rm w^{''}-c(-\kappa c)w=0}
\eqno(7b)
$$
and using Eq.~(6) one gets
$$
{\rm \left(\frac{d}{d\eta}+cu_{p}\right)
\left(\frac{d}{d\eta}-cu_{p}\right)w=
w^{''}-c(u_{p}^{'}+cu_{p}^{2})w=0}~.
\eqno(7c)
$$
To fix the ideas, we shall call Eq.~(7c) the bosonic equation.
On the other hand, the supersymmetric partner (or fermionic)
equation of Eq.~(7c) will be
$$
{\rm
\left(\frac{d}{d\eta}-cu_{p}\right)
\left(\frac{d}{d\eta}+cu_{p}\right)w_f=
w^{''}_{f}-c(-u_{p}^{'}+cu_{p}^2)w_{f}={\rm w^{''}_f
-c\cdot c_{\kappa, susy}w_f=0}}~.
\eqno(7d)
$$
Thus, one can write
$$
{\rm c_{\kappa,susy}(\eta)=-u_{p}^{'}+cu_{p}^2=
\left\{ \begin{array}{ll}
c(1+2{\rm tan}^2 c\eta) & \mbox{if $\kappa >1$}\\
c(-1+2{\rm coth}^2 c\eta) & \mbox{if $\kappa <1$}
\end{array} \right.}
$$
for the supersymmetric partner adiabatic index.
The solutions $\rm w_f$ are $\rm w_f =\frac{c}{\cos (c\eta +d)}$ 
and $\rm w_f =\frac{c}{sinh (c\eta)}$ for $\kappa =1$ and $\kappa =-1$,
respectively.

To find the general Riccati solution,
one is led to consider the following Riccati
equation ${\rm  cu_{g}^2 - \frac{d u_{g}}{d\eta}=cu^2_p -\frac{d u_p}{d\eta}}$,
whose general solution can be written down as 
${\rm u_{g}(\eta)= u_p(\eta) - \frac{1}{c}\frac{1}{v(\eta)}}$, 
where ${\rm v(\eta)}$ is 
an unknown
function. Using this ansatz, one obtains for the function ${\rm v(\eta)}$ the
following Bernoulli equation
$$
{\rm \frac{dv(\eta)}{d\eta} + 2c \, v(\eta)\, u_p(\eta) = 1},
\eqno(10)
$$
that has the solution
$$
{\rm v(\eta)= \frac{{\cal I}_{\kappa}(\eta)+ \lambda}{w_{\kappa}^{2}(\eta)}},
\eqno(11)
$$
where ${\rm {\cal I}_{\kappa}(\eta)= \int _{0}^{\eta} \,
w_{\kappa}^2(y)\, dy}$,
if we think of a half line problem for which $\lambda$ is a positive
integration constant thereby considered as a free SIDT parameter.

Thus, ${\rm u_{g}(\eta)}$ can be written as follows
$$
{\rm u_{g}(\eta;\lambda)=  u_p(\eta) -\frac{1}{c} \rm \frac{d}{d\eta}}
\Big[ {\rm ln}({\cal I}_{\kappa}(\eta) + \lambda) \Big]
\eqno(12a)
$$
$$
={\rm  \frac{d}{d\eta}
\Big[ ln \left(\frac{w_{\kappa}(\eta)}{{\cal I}_{\kappa}(\eta) +
\lambda}\right)^{\frac{1}{c}}\Big]}.
\eqno(12b)
$$
The important result provided by the SIDT is the one-parameter family of
adiabatic indices ${\rm c_{\kappa}(\eta;\lambda)}$
$$
{\rm -\kappa c_{\kappa}(\eta;\lambda)} = c{\rm u_{g}^2(\eta;\lambda) +
\frac{d u_{g}(\eta;\lambda)}{d\eta}}
\eqno(13a)
$$
$$
= {\rm -\kappa c - \frac{2}{c}
\frac{d^2}{d\eta^2} \Big[ ln({\cal I}_{\kappa}(\eta) + \lambda)}
\Big]
\eqno(13b)
$$
$$
= {\rm -\kappa c - \frac{4 w_{\kappa}(\eta) w_{\kappa}^{\prime}
(\eta)}{c({\cal I}_{\kappa}(\eta)
+ \lambda)}
+ \frac{2 w_{\kappa}^4(\eta)}{c({\cal I}_{\kappa}(\eta) + \lambda)^2}~.}
\eqno(13c)
$$
One can easily infer the following formula for the parametric adiabatic indices
$\gamma_{\kappa}(\eta;\lambda)$
$$
{\rm \gamma_{\kappa}(\eta;\lambda)=\gamma +\frac{8}{3\kappa(3\gamma-2)}
\frac{d^2}{d\eta^2} \Big[ ln({\cal I}_{\kappa}(\eta) + \lambda)
\Big]~,}
\eqno(13d)
$$
which I used for plotting.
All ${\rm  c_{\kappa}(\eta;\lambda)}$ have the same
supersymmetric partner index
${\rm c_{\kappa,susy}(\eta)
}$ obtained by deleting the zero mode
solution ${\rm w_{\kappa}}$.
They may be considered
as intermediates between the initial constant index ${\rm \kappa c}$ and
the supersymmetric partner
index ${\rm c_{\kappa,susy}(\eta)
}$.
From Eq.~($12b$) one can infer the new parametric `zero mode' solutions of
the universe for the family of barotropic
indices ${\rm c_{\kappa}(\eta;\lambda)}$ as follows
$$
{\rm w_{\kappa}(\eta;\lambda)= 
\frac{w_{\kappa}(\eta)}{{\cal I}_{\kappa}(\eta) + \lambda}
\Longrightarrow a_{\kappa}(\eta,\lambda)=
\left(\frac{w_{\kappa}(\eta)}{{\cal I}_{\kappa}(\eta) + \lambda}
\right)^{\frac{1}{c}}}~.
\eqno(14)
$$
A so-called double Darboux feature of the SIDT can be inferred by
writing the parametric family in terms of their unique ``fermionic" partner
index
$$
{\rm c_{\kappa} (\eta;\lambda)=c_{\kappa,susy}(\eta)
+\frac{2}{c}\frac{d^2}{d\eta^2}
\ln\left(\frac{1}{w_{\kappa}(\eta;\lambda)}\right)},
\eqno(15)
$$
which shows that the SIDT is of the
inverse Darboux type.\cite{MS} This is very important because it allows
a two-step (double Darboux) interpretation,
namely, in the first step one goes to the fermionic system and in the
second step one returns to a deformed bosonic system.
In addition, this interpretation helps me to introduce another
conjecture, on which I commented in a different context elsewhere.\cite{RNC}
In particle physics supersymmetric ideas forces one to introduce
supersymmetric partners for each known particle. However, if one promotes SIDT
at the level of a matter of principles, one can think of a scenario in
which the
supersymmetric partner system is unstable/unphysical and only the systems
generated by means of the general Riccati solution are encountered in
nature. Thus, in the cosmological context, I claim that only the
time-dependent barotropic fluids of indices $c_{\kappa}(\eta;\lambda)$ make
the real material content of the universe. At the cosmological scale,
the fluids of constant $\gamma$ occur in the limit
$\lambda \rightarrow \infty$ and therefore they look more as academic cases
in this context.
SIDT introduces a new time scale in zero-$\Lambda$ cosmology, which is
given by the integration constant $\lambda$. In ordinary supersymmetric
quantum mechanics, this constant is related to the contribution of the
irregular zero mode entering the SIDT parametric zero modes.
Extrapolating this fact to cosmology, one can say that this time constant
gives the contribution of the `unphysical' solution to the parametric
modes given by eq. (14). In the common approach, the `unphysical' solution
is discarded by imposing initial conditions.

Another comment I would like to make is that from the SIDT
standpoint I have worked at both fixed zero factorization `energy' and
fixed coupling constant ${\rm c}$ of the Schr\"odinger `potential'.
Thus, this is different from both standard quantum
mechanical framework, working at fixed coupling constant but variable energy,
and the Sturmian approach, which works at fixed energy but variable coupling
constant (depending only on quantum numbers). From the strictly
technical point of view,
what I have introduced here is a supersymmetric
class of solutions for the scale factor of the universe
corresponding to a family of cosmological fluids connected
to any given fluid of constant adiabatic index by the SIDT supersymmetric
procedure at zero factorization constant. One can interpret
the denominator in Eq.~(14) as an amplitude damping for these supersymmetric
scale factors whose origin is the time dependence of the adiabatic index.
Some plots for the matter-dominated universe (${\rm c=\frac{1}{2}}$) and
radiation-filled universe (${\rm c=1}$) are presented in Figs. 1(a)-(d) and
2(a)-(d), respectively.


\bigskip

{\bf 5} - In conclusion,
a supersymmetric class of cosmological fluids
has been introduced in this work possessing time-dependent adiabatic
indices. Such fluids may provide a simple
explanation for a currently accelerating universe and therefore
they should be of considerable interest. In future modeling
along the lines of this work the parameter $\lambda$ can be either
a constant fixed e.g., by some variational principles or can be made
dependent on other astroparticle parameters.
At the general level, a richer dynamics of the scale factor of the universe
is introduced by simple mathematical means without emphasizing
the phenomenological features, which are always of considerable speculative
origin. Moreover, the methods of this work can be easily applied to the 
majority of inflationary scenarios.

I would also like to point that in supersymmetry
breaking terms SIDT is of the unbroken type. Since several years, Witten
questions this fundamental issue.\cite{W94} Most recently, he
remarked:\cite{W00}
``In the standard framework of low energy physics,
{\em there appears to be no
natural explanation for vanishing or extreme smallness of the vacuum
energy}, while on the other hand it is very difficult to modify this
framework in a sensible way. In seeking to resolve this problem, {\em one
naturally wonders if the real world can somehow be interpreted in terms of
a vacuum state with unbroken supersymmetry}". In the cosmological context
of this work,
{\em there appears to be a simple technical explanation for the
vanishing of the vacuum energy}, because only in this case Eq.~(4)
reduces to a Riccati equation wherefrom one can proceed with SIDT, which is
naturally of unbroken type and moreover
it might be used to explain the supernovae Ia data as I argued herein.
In the vacuum case, the supersymmetric one-parameter adiabatic indices are
${\rm 
\gamma_{\kappa,vac}(\eta;\lambda)=-\frac{4}{3\kappa}
\frac{d^2}{d\eta^2} \Big[ ln({\cal I}_{\kappa}(c\eta) + \lambda)
\Big]}$, leading to plots similar to the other parametric barotropic fluids
(See Figs.~3(a) and (b)).

Finally, I mention that in the framework of supersymmetric quantum mechanics
I published a direct iteration of this type of SIDT leading
to multiple parameter `zero modes' \cite{RZM}, which can be adapted to
cosmology if the approach presented here is proven relevant.

\nonumsection{Acknowledgements}

\noindent
This work was supported by a Project from CONACyT (No.~458100-5-25844E).

\end{document}


\newpage

\runninghead{Haret C. Rosu} 
{Negative friction}

\normalsize\textlineskip
\thispagestyle{empty}
\setcounter{page}{1}


\vspace*{0.88truein}

\bigskip
\centerline{\bf  NEGATIVE FRICTION OF DARBOUX COSMOLOGICAL FLUIDS}  
\vspace*{0.035truein}
\vspace*{0.37truein}
\vspace*{10pt}
\centerline{\footnotesize H.C. ROSU}
\vspace*{0.015truein}
\centerline{\footnotesize  Instituto de F\'{\i}sica,
Universidad de Guanajuato, Apdo Postal E-143, 37150 Le\'on, Gto, Mexico}
\vspace*{0.225truein}

\vspace*{0.21truein}
\abstracts{Atmospheric free fall with quadratic resistance when 
considered from the point of view of the strictly isospectral 
construction of supersymmetric quantum mechanics is
a powerful analogy for the class of 
Darboux cosmological fluids that I introduced in a previous work.
Here, using this analogy, 
it is shown that the possible accelerating cosmological
effects are due to the quadratic negative friction
properties with which the Darboux fluids are endowed through the 
mathematical construction.}{}{}


\textlineskip                  
\vspace*{12pt}                 

\vspace*{1pt}\textlineskip	
\vspace*{-0.5pt}
\noindent


\noindent





{\bf 1} - It has recently been realized by Faraoni,\cite{Far} that the equations 
describing barotropic FRW cosmologies can be combined in a simple 
Riccati equation leading in a much easier way 
to the textbook solutions for the FRW scale factors.
In a previous work, using mathematical schemes belonging to 
supersymmetric quantum mechanics, I have employed the FRW Riccati equation to
introduce a so-called Darboux class of cosmological fluids with time-dependent
adiabatic indices $\gamma (\eta;\lambda)$.\cite{Ros}  
This type of fluids could provide an explanation for the     
recent data from high-redshift Type Ia (hydrogen-deficient)
supernovae indicating
that the universe has {\em now} an accelerating 
large scale expansion.\cite{type1a} In this note, I would like to clarify
the nature of the effects produced by the mathematical scheme through  
the close analogy with the atmospheric resistive free fall.  

\bigskip

{\bf 2} - Faraoni obtained the following cosmological Riccati equation
$$
 \frac{du}{d\eta}+cu^2+\kappa c=0
\eqno(1)
$$
for the log derivative of the FRW scale factor $u=\frac{da/d\eta
}{a}$.
The independent variable is the conformal time $\eta$, 
$c=\frac{3}{2}\gamma -1$ is assumed positive, 
and $\kappa=0,\pm1$ is the curvature index of the flat, closed, open
FRW universe, respectively.

From the mathematical point of view, the FRW Riccati equation is of constant 
coefficients and therefore is directly integrable. In general, 
for a Riccati equation of the type 
$$
 \frac{dy}{dx}=\alpha y^2+\beta~,
\eqno(2)
$$
the solutions can be written as follows

\bigskip

(i) ${\rm for} \quad  \frac{\alpha}{\beta}>0$
$$
y=\sqrt{\frac{\beta}{\alpha}}\tan\left
(\sqrt{\frac{\alpha}{\beta}}\beta x+\phi _1\right),
\qquad {\rm or} \qquad 
y=\sqrt{\frac{\beta}{\alpha}}\cot
\left(-\sqrt{\frac{\alpha}{\beta}}\beta x+\phi _2\right)
\eqno(3)
$$

\bigskip

(ii) ${\rm for} \quad \frac{\alpha}{\beta}< 0$
$$
y=\sqrt{-\frac{\beta}{\alpha}}\tanh
\left(\sqrt{-\frac{\alpha}{\beta}}\beta x+
\varphi _1\right),\quad {\rm or} \quad 
y=\sqrt{-\frac{\beta}{\alpha}}\coth\left(\sqrt{-\frac{\alpha}{\beta}}\beta x+
\varphi _2\right)~.
\eqno(4)
$$
One can see that the cosmological cases correspond to
$$
\alpha  =-c, \qquad \beta  =-\kappa c \qquad 
\Longrightarrow \quad \frac{\alpha}{\beta}=
\frac{1}{\kappa}~,
\eqno(5)
$$
and therefore the two mathematical cases are directly connected to the type of 
FRW universe. The case $\kappa =0$ is presented in [\refcite{HCR}].

The time-dependent barotropic indices $\gamma _{\kappa} (\eta; \lambda )$
have been introduced through the 
so-called strictly isospectral Darboux technique (SIDT). The latter can be 
interpreted as a three-step procedure in which one goes from a
physical Riccati problem to an
unphysical one by changing the free term and the 
sign of the derivative term and then comes
back to another physical problem similar to the original one by the same 
type of changes. However, in the 
third-step problem, unphysical effects of the intermediate problem are 
maintained as a consequence of the applied scheme, essentially because the 
general Riccati solution of the intermediate problem is  
used in the last step as well. The very illustrative example 
of the classical Newton motion with quadratic resistance is presented 
below since it both helps understanding the three-step character of SIDT
and provides further insight in the nature of Darboux cosmological fluids.

\bigskip

{\bf 3} - 
Newton's law of force for classical motions with quadratic resistance, such as
gravitational free fall in terrestrial atmosphere and similar motion of 
electric charges in constant electric fields, 
is Riccati's equation without linear term. 
About 40 years ago, the free fall in quadratic resistive atmosphere
has been studied in some detail by Davies in his book.\cite{D}
The equation of motion is of the following type

$$
m\frac{d^2s}{dt^2}=mg-K\left(\frac{ds}{dt}\right)^2~,
\eqno(6)
$$
which can be written
$$
\frac{dv}{dt}=g-\epsilon v^2~,
\eqno(7)
$$
where $\epsilon=\frac{K}{m}$ is the friction coefficient per mass unit. 
Again, comparing this Newtonian case with Eq.~(2) one gets
$$
\alpha =-\epsilon, \qquad \beta =g \qquad \Longrightarrow \quad
\frac{\alpha}{\beta}=
\frac{-\epsilon}{g}<0,
\eqno(8)
$$
and therefore the solution is given by
$$
v(t)=\sqrt{\frac{g}{\epsilon}}\tanh\left(\sqrt{\epsilon g}t+
\varphi _1\right)~,
\eqno(9)
$$
which is Davis' result. The falling velocity $v(t)$ presents a substantial
time dependence - acceleration stage - only for a short period and rapidly 
turns into the limiting constant velocity $V=\sqrt{\frac{g}{\epsilon}}$.
The phase $\varphi _1$ is related to the initial velocity. It is not important 
for the general discussion here and therefore we take $\varphi _1=0$.  

Let us sketch now shortly, the three steps of SIDT for the resistive 
atmospheric free fall:

(i) The initial equation is $\frac{dv}{dt}=g-\epsilon v^2$ representing the 
normal resistive falling motion.
  
(ii)
One shifts to the supersymmetric `unphysical' partner motion of the form
$$
\frac{-dv}{dt}=g_2(t)- \epsilon v^2~, 
\eqno(10)
$$
where
$$
g_2(t)=g[-1+2\tanh ^2(\sqrt{\epsilon g}t)]~.
\eqno(11)
$$

The negative sign in Eq.~(10) shows that the motion is upwards and  
through a fluid medium endowed with friction of 
negative friction coefficient (i.e., {\em antifriction}) as the driving
force, plus another force $g_2(t)$ of driving character for 
$t<T=\frac{1}{\sqrt{\epsilon g}}{\rm Arctanh} 
\frac{1}{\sqrt{2}}$
and turning friction-like afterwards.

The first two steps are connected to each other through the particular 
solution of the initial normal motion as given by Eq.~(9). 

(iii)
The third step is the return back 
to an equation similar to the initial one, namely
$$
\frac{dv}{dt}=g_3(t; \lambda)- \epsilon v^2~, 
\eqno(12)
$$
where 
the time-dependent forces $g_3(t;\lambda)$ are given by
$$
g_3(t;\lambda)=g\Bigg[1-2\frac{d^2}{dt^2}
\ln(I_{b}(t) +\lambda)\Bigg]~,
\eqno(13)
$$ 
where $I_b(t)=\int _{0}^{t}{\rm cosh}^2xdx$.

The second and the third steps are also connected to each other through the 
general solution of the `unphysical' problem. Thus, although the motion given
by Eq.~(12) looks normal, its solution $v(t)$ corresponds to 
a rising on the vertical rather than a fall. Exactly as in the falling case,
the rising has a short accelerating stage followed by a uniform rising with
constant limiting velocity. Indeed, for sufficiently large times,
the driving antifriction force is 
at its maximum value $-\epsilon V^2$, whereas $g_2(t\rightarrow \infty)
=+g$ is a constant friction force and one gets the value of $V$ from
the balance of the two forces.   


\bigskip



\bigskip

{\bf 4} - Eqs.~(5) and (8) give the translation rules
for the analogy between the cosmological case 
and the atmospheric friction case. 
From Eq.~(1) one sees that the normal cosmological friction force is $cu^2$.
The equivalent of the upward velocity is 
$u_{\kappa}=\frac{1}{a_{\kappa}(\eta;\lambda)}\frac{da_{\kappa}(\eta;\lambda)}
{d\eta}$, whereas the ones for $g_3(t;\lambda)$ are the time-dependent indices
$\gamma _{\kappa} (\eta; \lambda )$.
The analogy also predicts that the accelerating period is limited in time. 
In this `quadratic friction' picture, the cosmological SIDT as worked out in
our main paper [\refcite{Ros}] is of unphysical nature as far as the `antifriction' properties
of the Darboux fluids should be invoked through the second step of SIDT.
However, cosmology is perhaps the only branch of science where  
unphysical effects are manifest. The aforementioned
SIDT atmospheric rising looks completely unphysical but the expansion of the 
universe is a reality.  


\nonumsection{Acknowledgements}

\noindent
The author thanks Dr. M. Nowakowski and F. Aceves de la Cruz for discussions.



\runninghead{H.C. Rosu $\ldots$} {Errata and Addenda $\ldots$}


\normalsize\textlineskip
\thispagestyle{empty}
\setcounter{page}{1}

\copyrightheading{}			

\vspace*{0.88truein}

\centerline{\bf ERRATA AND ADDENDA}
\bigskip
\bigskip
\begin{center}
{\bf DARBOUX CLASS OF COSMOLOGICAL FLUIDS WITH TIME-DEPENDENT ADIABATIC 
INDICES}
\end{center}
\vspace*{0.035truein}
\vspace*{0.37truein}
\centerline{\footnotesize H.C. Rosu}
\centerline{[Mod. Phys. Lett. A, Vol. 15, No. 15 (2000) 979-990]}
\vspace*{0.015truein}
\baselineskip=10pt
\vspace*{10pt}
\vspace*{0.225truein}

\vspace*{0.21truein}
\abstracts{
A number of corrections to the work in which I introduced one-parameter
Darboux supersymmetric barotropic fluids in cosmology are performed. 
As addenda, the flat FRW case, which was left apart, is presented in detail
in the supersymmetric approach. \\
}{}{}


\textlineskip                  
\vspace*{12pt}                 

\vspace*{1pt}\textlineskip	
\vspace*{-0.5pt}
\noindent


\noindent




\noindent






\noindent
{\bf Errata}. 

\noindent
On p. 981

(i) In the second line of the last paragraph the tanh function should be coth.

\noindent
On p. 982 

(ii) The function $\rm w$ in Equation (7d) should carry the 
subindex $\rm f$ to 
emphasize that the solution $\rm w_f$ is not the same as in Eq.~(7c), i.e., 
$$
{\rm
\left(\frac{d}{d\eta}-cu_{p}\right)
\left(\frac{d}{d\eta}+cu_{p}\right)w_f=
w^{''}_{f}-c(-u_{p}^{'}+cu_{p}^2)w_{f}={\rm w^{''}_{f}
-c\cdot c_{\kappa, susy}w_f=0}}~,
\eqno(7d)
$$

(iii) In the unnumbered equation following (7d) the $\rm tan$ function 
should be squared, whereas the $\rm tanh$ should be coth squared.

(iv) At the end of the same paragraph, the following sentence should be 
included:
The solutions $\rm w_f$ are $\rm w_f =\frac{c}{\cos (c\eta +d)}$ 
and $\rm w_f =\frac{c}{sinh (c\eta)}$ for $\kappa =1$ and $\kappa =-1$,
respectively. 

(v) The general Riccati solution should be written $u_g(\eta)=u_p (\eta)-
\frac{1}{{\rm c}}\frac{1}{v(\eta)}$. This leads to a free term equal to unity
in the Bernoulli equation (10) and to the dropping of $\rm c$ everywhere in the 
argument of the integral ${\cal I}$. In addition, in Eq.~(11) $\rm c$ in the
argument of $w_{\kappa}^{2}$ should be dropped.

(vi) 
In Eq.~(12b), ${\cal I}_{c\kappa}(\eta)$ should be replaced by 
${\cal I}_{\kappa}(\eta)$.

(vii) 
The left hand side of Eq.~(13a) should be $\rm -\kappa c_{\kappa}(\eta;\lambda)$.

\noindent
On p. 990, the page number of Ref. 8 is 4850 and not 2850, whereas the 
e-print at Ref. 15 should be hep-ph/0002297.

\bigskip
\newpage

\noindent
{\bf Addenda}.

\noindent
We would like to present shortly the flat case $\kappa =0$ in the 
supersymmetric approach. 
For this case the analog of Eq.~(4) at page 980 reads
$$
{\rm
\frac{\ddot{a}}{a}+c\left(\frac{\dot{a}}{a}\right)^2=0~,}
\eqno(1)
$$
where ${\rm c=\frac{3}{2}\gamma -1}$. 
Writing $\rm b=\frac{\dot{a}}{a}$,
Eq.~(1) is rewritten in the form
$$
\rm \dot{b}+(c+1)b^2=0~.
\eqno(2)
$$
There are two cases.

(i) For $\rm c+1\neq 0$, the solution is $\rm b _p =\frac{1}{c+1}\frac{1}{t}$, or in 
terms of the scale factor 
$\rm a =a_0 t^{\frac{1}{c+1}}=a_0t^{\frac{2}{3\gamma}}$, which is a standard 
result.

(ii) For $\rm c+1=0$ ($\gamma =0$), one gets $\rm \dot{b}=0$, 
i.e., 
$\rm b_p=\frac{\dot{a}}{a}=$ const =$H$ (in cosmology). 
Thus, we get the standard result $\rm a =a_0\exp (Ht)$. This case is completely
degenerate from the point of view of the supersymmetric constructions. 

\bigskip


For the first case, the solutions $\rm b _p$ enter factorizations of the following type
$$
{\rm \left(\frac{d}{dt}+(c+1)b_{p}\right)
\left(\frac{d}{dt}-(c+1)b_{p}\right)w=
\ddot{w}-(c+1)[\dot{b}_{p}+(c+1)b_{p}^{2}]w=0}~.
\eqno(3)
$$
Thus, because of Eq.~(2), we get $\rm \ddot{w}=0$.
The solution is a linear function $\rm w_{b} \propto t$.
The supersymmetric partner (or fermionic)
equation with respect to Eq.~(3) will be
$$
{\rm
\left(\frac{d}{dt}-(c+1)b_{p}\right)
\left(\frac{d}{dt}+(c+1)b_{p}\right)w_f=
\ddot{w}_{f}-(c+1)[-\dot{b}_{p}+(c+1)b_{p}^2]w_{f}=0}~.
\eqno(4)
$$
This leads to ${\rm \ddot{w}_{f}-\frac{2}{t^2}w_{f}=0}$. The linearly
independent solutions are $\rm w_{f1}=t^2$ and $\rm w_{f2}=t^{-1}$.  

The general Riccati solution comes out from 
$$
{\rm  (c+1)b_{g}^2 - \frac{d b_{g}}{dt}=(c+1)b^2_p -\frac{d b_p}{dt}}~,
$$
whose general solution can be written down as 
${\rm b_{g}(t)= b_p(t) - \frac{1}{c+1}\frac{1}{v(t)}}$, where ${\rm v(t)}$ is 
an unknown
function. Using this ansatz, one obtains for the function ${\rm v(t)}$ the
following Bernoulli equation
$$
{\rm \frac{dv(t)}{dt} + 2(c+1) \, v(t)\, b_p(t) =1},
\eqno(5)
$$
that has the solution
$$
{\rm v(t)= \frac{{\cal I}_{0}(t)+ \lambda}{w_{b}^{2}(t)}}~,
\eqno(6)
$$
where ${\rm {\cal I}_{0}(t)= \int _{0}^{t} \,
w_{b}^2(y)\, dy\propto t^3/3}$,
and we consider $\lambda$ as a positive integration constant that is employed as 
a free parameter.

Thus, ${\rm b_{g}(t)}$ can be written as follows
$$
{\rm b_{g}(t;\lambda)=  b_p(t) -
\rm \frac{1}{c+1}\frac{d}{dt}}
\Big[ {\rm ln}({\cal I}_{0}(t) + \lambda) \Big]
\eqno(7)
$$
$$
={\rm  \frac{d}{dt}
\Big[ ln \left(\frac{w_{b}(t)}{[{\cal I}_{0}(t) +
\lambda]}\right)^{\frac{1}{c+1}}\Big]}.
\eqno(8)
$$
One can get a one-parameter family of quantities ${\rm P_{b}(t;\lambda)}$
as follows
$$
{\rm  P_{b}(t;\lambda)} = (c+1){\rm b_{g}^2(t;\lambda) +
\frac{d b_{g}(t;\lambda)}{dt}}
\eqno(9)
$$
$$
= {\rm  - \frac{2}{c+1}
\frac{d^2}{dt^2} \Big[ ln({\cal I}_{0}(t) + \lambda)}
\Big]
\eqno(10)
$$
$$
= {\rm - \frac{4 w_{b}(t) \dot{w}_{b}
(t)}{(c+1)({\cal I}_{0}(t)
+ \lambda)}
+ \frac{2 w_{b}^4(t)}{(c+1)({\cal I}_{0}(t) + \lambda)^2}~.}
\eqno(11)
$$
The interpretation of $\rm P_b$ is similar to that in the original paper, namely as 
the general Riccati contribution to any constant barotropic index $\gamma$.
The corresponding  parametric scale factors of the universe 
will be
$$
{\rm 
a_{b}(t,\lambda)=\left(\frac{w_{b}(t)}{[{\cal I}_{0}(t) + \lambda]}
\right)^{\frac{1}{c+1}}}~.
\eqno(12)
$$

Plots of $\rm \gamma (t;\lambda)$ and $\rm a_{b}(t;\lambda)$
for the dust universe and for a particular value of $\lambda$
are presented in Figs.~1 and 2, respectively, together with the standard 
counterparts. 
\bigskip












\vskip 2ex
\centerline{
\epsfxsize=180pt
\epsfbox{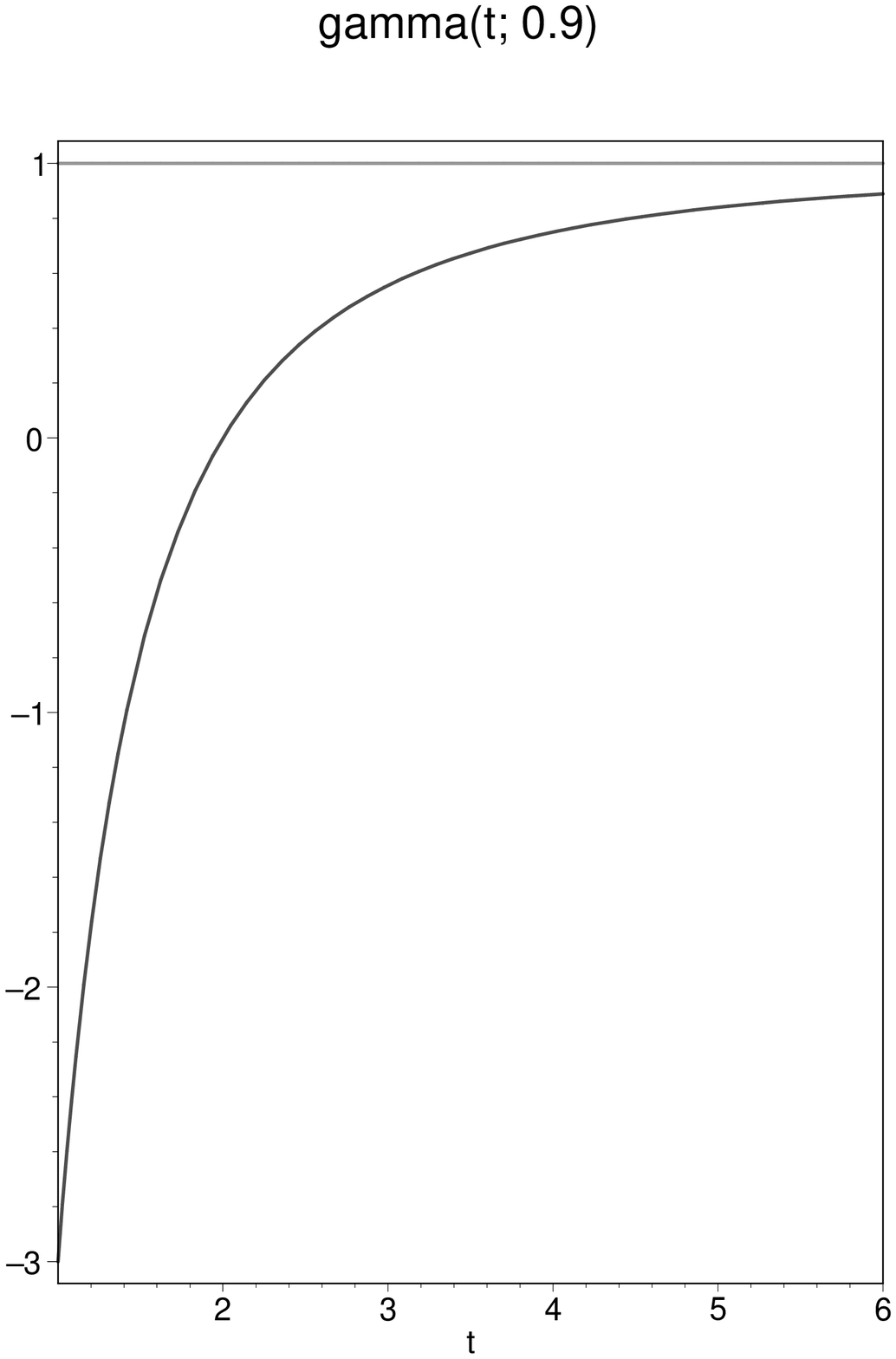}}
\vskip 4ex
\begin{center}
{\small{Fig. 1}$\quad$
Time-dependent adiabatic index of a Darboux barotropic fluid of 
dust type for $\lambda=0.9$. There is a $-1/t^2$ singularity at the origin,
after which the index becomes rapidly $\gamma =1$.}
\end{center}

\vskip 2ex
\centerline{
\epsfxsize=180pt
\epsfbox{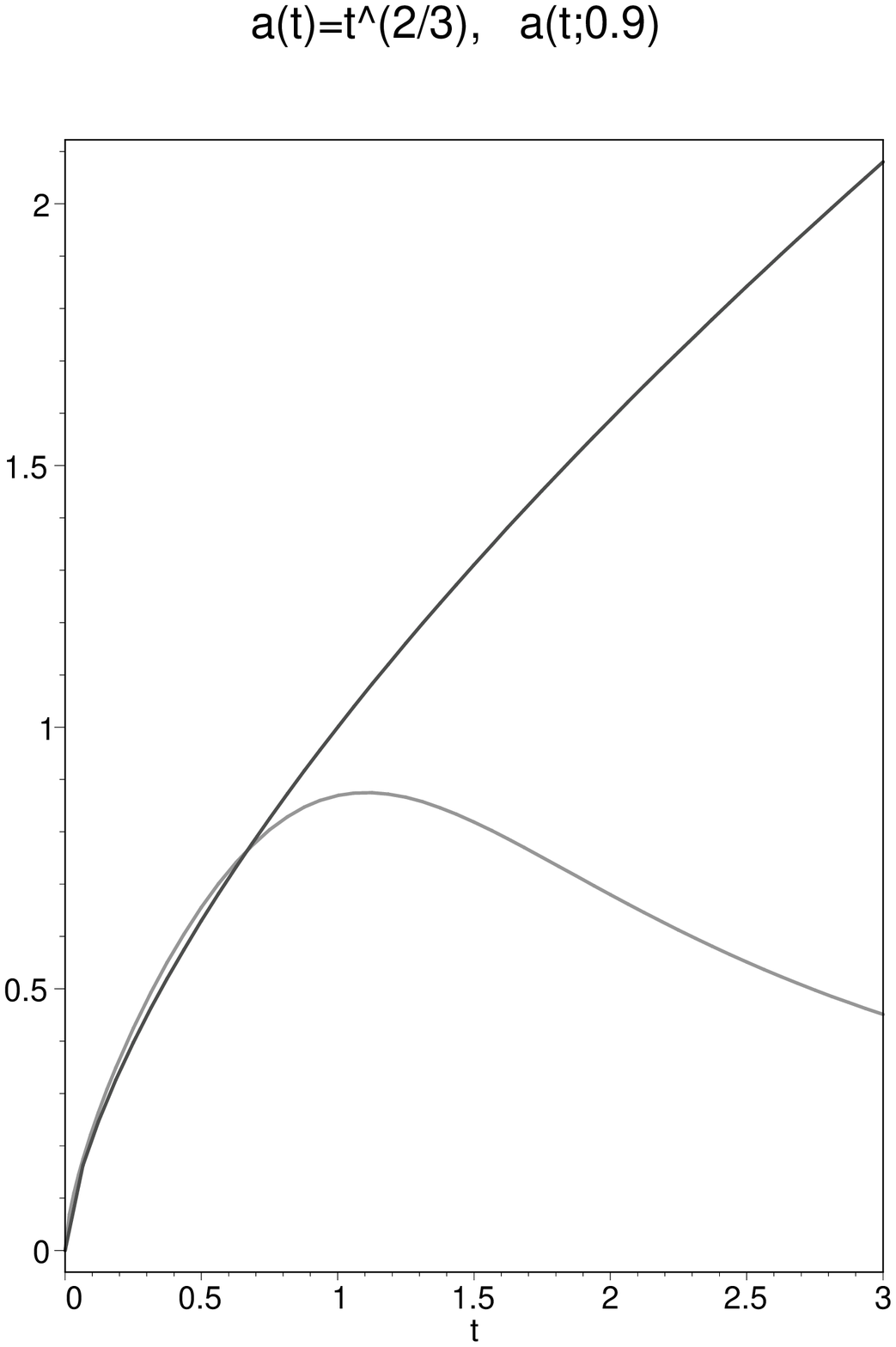}}
\vskip 4ex
\begin{center}
{\small{Fig. 2}$\quad$
Scale factors of a $\gamma =1$ barotropic flat universe and of the corresponding
one-parameter supersymmetric flat universe (for $\lambda=0.9$).}
\end{center}